\documentclass[preprint,superscriptaddress,tightenlines,preprintnumbers,showpacs,showkeys,nofootinbib]{revtex4}        
\usepackage{graphicx}
\usepackage{dcolumn}
\usepackage{bm}
\newcommand{\be}{\begin{equation}}
\newcommand{\ee}{\end{equation}}
\newcommand{\bee}{\begin{eqnarray}}
\newcommand{\eee}{\end{eqnarray}}
\newcommand{\eq}{\end{quote}}
\newcommand{\nn}{\nonumber}
\newcommand{\Slash}[1]{\ooalign{\hfil/\hfil\crcr$#1$}}

\begin{document}      

\preprint{\vbox{\hbox{PNU-NTG-04/2004}\hbox{PNU-NURI-03/2004}}}
\title{Photo-production of the pentaquark $\Theta^{+}$ with positive
and negative parities}
\author{Seung-Il Nam}
\email{sinam@rcnp.osaka-u.ac.jp}
\affiliation{Research Center for Nuclear Physics (RCNP), Ibaraki,
Osaka 567-0047, Japan}
\affiliation{Department of
Physics and Nuclear physics \& Radiation technology Institute (NuRI),
Pusan National University, Busan 609-735, Korea}  

\author{Atsushi Hosaka}
\email{hosaka@rcnp.osaka-u.ac.jp}
\affiliation{Research Center for Nuclear Physics (RCNP), Ibaraki, Osaka
567-0047, Japan}

\author{Hyun-Chul Kim}
\email{hchkim@pusan.ac.kr}
\affiliation{Department of
Physics and Nuclear physics \& Radiation technology Institute (NuRI),
Pusan National University, Busan 609-735, Korea} 
\date{February 16, 2004}

\begin{abstract}
We investigate the production of the pentaquark $\Theta^+$ baryon via the
$\gamma n\rightarrow  K^-\Theta^+$ and $\gamma p\rightarrow
\overline{K}^0 \Theta^+$ 
processes, focusing on the parity of the $\Theta^+$.  Using the
effective Lagrangians, we calculate the total and differential 
cross sections with the spin of the $\Theta^+$ presumed to be $1/2$.
We employ the coupling constant of the $KN\Theta$ vertex determined by
assuming its mass and the decay width to be $1540\,{\rm MeV}$ and 
$15\,{\rm MeV}$.  That of the $K^*N\Theta$ is taken to be about a
half of the $KN\Theta$ coupling constant.  We estimate the cutoff parameter
by reproducing the total cross section of the $\gamma p \rightarrow
K^+ \Lambda$ reaction.  It turns out that the total cross section for
the $\gamma n\rightarrow K^- \Theta^+$ process is about four times
larger than that of the $\gamma p\rightarrow\overline{K}^0\Theta^+$.  
We also 
find that the cross sections for the production of the positive-parity 
$\Theta$ are about ten times as large as those for the negative-parity
ones.     
\end{abstract}

\pacs{13.60.Le, 13.75.Jz, 13.85.Fb}
\keywords{Pentaquark baryon, Photo-production of the
  $\Theta^+$, Parity of the $\Theta^+$}
      
\maketitle

\section{introduction}
Since the discovery of the pentaquark $\Theta^+$ baryon by
the LEPS collaboration~\cite{Nakano:2003qx}, motivated by 
the theoretical work by Diakonov
{\em et al.}~\cite{Diakonov:1997mm}, the physics of the pentaquark
states has become a very hot issue in hadron physics.  
The subsequent experiments confirmed the existence of the
$\Theta^+$~\cite{Barmin:2003vv,Stepanyan:2003qr,Kubarovsky:2003fi,Barth:2003es,Airapetian:2003ri}.
Its mass is around 1530 MeV with   
still about 20 MeV uncertainty, whereas only 
the upper bound is established for its width 
($<25\,{\rm MeV}$).  
However, considering the fact that the DIANA collaboration
has reported that its decay could be remarkably narrow 
($< 9\, {\rm MeV}$)~\cite{Barmin:2003vv}, where the 
experimental energy resolution was significantly smaller than 
this number~\cite{private1}, the narrowness of the width 
is likely a characteristic
of the pentaquark $\Theta^+$.  The NA49 
collaboration~\cite{Alt:2003vb} has announced another exotic
pentaquark baryon $\Xi_{3/2}$, the width of which is also very   
narrow.  
Prasza\l owicz~\cite{Praszalowicz:2003tc} has pointed out that
the smallness of the width can be explained in the large 
$N_c$ limit with SU(3) symmetry breaking.  
Karliner and Lipkin suggested an explanation based on a model with two
diquarks and one antiquark~\cite{Karliner:2004qw}.     

It is also of great importance to determine the quantum numbers of the
$\Theta^+$.  Since the $\Theta^+$ decays into a neutron and a $K^+$,
its strangeness is determined to be  $S=+1$.  Its isospin $T=0$ has
been inferred from the SAPHIR~\cite{Barth:2003es} and HERMES 
collaborations~\cite{Airapetian:2003ri} which have found no signal of the
$\Theta^{++}$.  On the other hand, the spin and parity of the
$\Theta^+$ are still not known to date experimentally, which brought
about a great deal of theoretical works to focus on determining its
parity.  However, there has been no agreement on its parity 
at all.  While the chiral soliton model~\cite{Diakonov:1997mm} prefers
the positive parity, QCD sum
rules~\cite{Sugiyama:2003zk,Zhu:2003ba} predict its parity to  
be negative.   The lattice QCD~\cite{Sasaki:2003gi,Csikor:2003ng} also
supports the negative parity.  In the chiral constituent quark 
model~\cite{Stancu:2003if,Glozman:2003sy} with the spin-flavor
interaction and in the chiral bag model~\cite{Hosaka:2003jv}, 
the positive-parity state turns out to be more stable 
than the negative-parity one due to the interaction inspired by 
chiral symmetry.  
However, Huang {\em at al.}~\cite{Huang:2003we} argue that 
if $u$(or $d$) - $\overline{s}$
interaction is considered, the negative-parity state produces the
$\Theta^+$ mass closer to the experimental value.    

A great amount of investigation on the production of the pentaquark
baryons via various processes has been already performed.  Its
hadron-induced production also has been investigated in  
Refs.~\cite{Liu:2003rh,Hyodo:2003th,Oh:2003kw,nam2}.  In particular,
Refs.~\cite{nam2,nam3} have scrutinized the parity of
the $\Theta^+$ in its production via the $NN$ interaction,
motivated by a series of recent
works~\cite{Thomas:2003ak,Hanhart:2003xp}.  It was  
found that the cross sections for the production of the
positive-parity $\Theta^+$ are approximately ten times larger than
those for the negative-parity ones.  
The photo-production of the
$\Theta^+$ has been also studied in the Born
approximation~\cite{Liu:2003zi,Nam:2003uf,Oh:2003kw,Zhao:2003gs,
Yu:2003eq,Liu:2003ab,Li:2003cb,Ko:2003xx}.    

In the present work, we would like to extend our former study of the
$\gamma n\rightarrow K^{-}\Theta^{+}$ and 
$\gamma p\rightarrow\overline{K}^{0}\Theta^{+}$ 
reactions~\cite{Nam:2003uf}.  
We attempt to provide physical interpretation for the obtained results 
whenever possible and extract items which we can discuss in a
model-independent manner.  In the present work we investigate rather
carefully the role of the vector meson $K^*(892)$ which was not
included in the previous work~\cite{Nam:2003uf}.  Since the vector
meson $K^*$ plays an important role in the $\gamma p\rightarrow
{K}^{0}\Lambda$, it is expected to be so also for the $\Theta^+$
production.  While the coupling constants of $K$ exchange can be
determined by using the width and mass of the $\Theta^+$, we do not
have any information of that of $K^*$ exchange.  Hence, we will follow
Ref.~\cite{Liu:2003zi} in which the value of the coupling constant for
the $K^* N\Theta$ vertex is chosen to be about a half of that for the
$KN\Theta$, reasoning that the empirical value of the $KN\Lambda$
($KN\Sigma$) is approximately twice as large as that of the $K^*
N\Lambda$ ($K^*N\Sigma$).  In order to calculate the cross sections of
the $\Theta^+$ photo-production, its magnetic moment also has to be 
considered.  Due to the lack of experimental information  
on the electro-magnetic structure of the pentaquark states, one has to
rely on model calculations to determine its magnetic moment.  The
magnetic moment of the $\Theta^+$ has been already
estimated in various models~\cite{Nam:2003uf,Zhao:2003gs,Kim:2003ay,
Huang:2003bu,Liu:2003ab}.  Its value varies in the range
$0.1\sim0.3\,\mu_N$, where $\mu_N$ is the nuclear magneton.  Thus, we
will use in this work the anomalous magnetic moment
$\kappa_\Theta=-0.8$.   

In order to take into account the extended size of hadrons, it is
essential to introduce a form factor at each vertex.  However, its
presence violates the gauge invariance of the electromagnetic
interaction.  It is due to the fact that the form factors bring about
the non-locality in the interaction~\cite{Ohta:ji}.  Hence, we need to
restore the gauge invariance.  While there is no theoretical firm
ground to remedy this gauge-invariance problem caused by form factors,
various Refs.~\cite{Ohta:ji,Haberzettl:1998eq,Davidson:2001qs} put
forward several 
prescriptions for the form factors to restore the gauge 
invariance.  In this work, we closely follow the method suggested by
Ref.~\cite{Davidson:2001qs}.  In addition, we estimate the cutoff parameters by
reproducing the total cross sections for the photo-production
of the $\Lambda$.  

This paper is organized as follows: In Section II, we will
describe a method to calculate the Feynman invariant amplitude for the
processes 
$\gamma n\rightarrow \Theta^+ K^-$ and 
$\gamma p\rightarrow \overline{K}^0\Theta^+$.  
We  will also discuss the gauge-invariant form
factor and two different schemes of the pseudovector (PV) and 
pseudoscalar (PS) couplings.  In the subsequent section, we will
present the numerical results for the total and differential cross
sections for the two different parities of the $\Theta^{+}$ and will
discuss them in comparison with other models.  In Section IV we will
summarize the results and draw a conclusion.  

\section{General formalism}
\begin{figure}[tbh]
\begin{tabular}{c}
\resizebox{15cm}{3cm}{\includegraphics{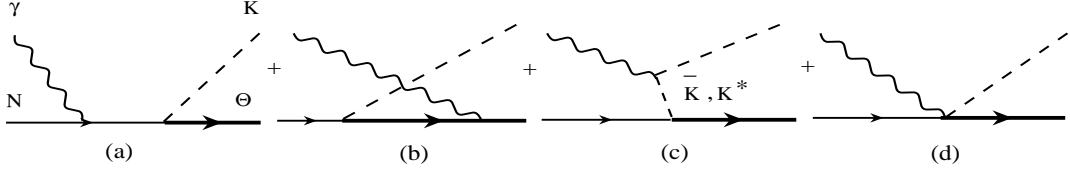}}
\end{tabular}
\caption{Diagrams for the photo-production of the $\Theta^+$.}
\label{dia}
\end{figure}

Relevant diagrams for the photo-production of the $\Theta^+$ are drawn
in Fig.~\ref{dia}.
Concerning the $KN\Theta$ vertex, we utilize two different
interactions, {\em i.e.}, the pseudoscalar (PS) and pseudovector (PV)
schemes.  The effective Lagrangians for the reactions are 
given as follows:
\bee
\mathcal{L}_{N\Theta K}
&=&
ig\overline{\Theta}\Gamma_{5}K N + (\mbox{h.c.}),  
\nn\\
\mathcal{L}_{N\Theta K}
&=&
-\frac{g^{\ast}_{A}}{2f_{\pi}}
\overline{\Theta}\gamma_{\mu}\Gamma_{5}\partial^{\mu}K N + ({\rm h.c.}),
\label{ntkpv} \nn\\
\mathcal{L}_{\gamma KK}
&=&
ie \left\{ K (\partial^{\mu}\overline{K}) -
  (\partial^{\mu}K)\overline{K}\right\} A_{\mu} + ({\rm h.c.}),
\nn\\
\mathcal{L}_{\gamma NN}
&=&
-e\overline{N}\left(\gamma_{\mu}
+i\frac{\kappa_{N}}{2M_{N}}
\sigma_{\mu\nu}k^{\nu}\right)N \, A^{\mu} + ({\rm h.c.}), 
\nn\\
\mathcal{L}_{\gamma\Theta \Theta}
&=&
-e\overline{\Theta}\left(\gamma_{\mu}
+i\frac{\kappa_{\Theta}}{2M_{\Theta}}
\sigma_{\mu\nu}k^{\nu}\right)\Theta \, A^{\mu}\ + ({\rm h.c.}), 
\label{gtt}
\eee 
where $\Theta$, $N$, and $K$ stand for the pentaquark 
$\Theta^+$, the nucleon, and the kaon fields, respectively.  
Parameters $e$,
$\kappa$, and $M$ designate the electric charge, the anomalous
magnetic moment, and the mass of baryon, respectively.  $\Gamma_{5}$
is generically $\gamma_{5}$ for the positive-parity $\Theta^{+}$
($\Theta_{+}^{+}$) and ${\bf 1}_{4\times 4}$ for the negative-parity
$\Theta^{+}$ ($\Theta_{-}^{+}$).  In the case of the positive-parity
$\Theta^+$, the coupling constants for $K$
exchange can be determined by using the decay width
$\Gamma_{\Theta\rightarrow KN} = 15\, {\rm MeV}$ and the mass
$M_{\Theta} = 1540\, {\rm MeV}$, from which we obtain 
$g^{\ast}_{A}=0.28$ for the PV interaction as well as $g=3.8$ for the
PS.  Similarly, we find $g^{\ast}_{A}=0.16$ and $g=0.53$ for the
negative-parity one. 

$K^{*}$ exchange is also taken into account in this work as in
Refs.~\cite{Liu:2003rh,Liu:2003zi,Oh:2003kw,Janssen:2001wk}.  The
corresponding Lagrangians are given as follows:
\bee
\mathcal{L}_{\gamma {K} K^{*}}&=& g_{\gamma K K^{*}}
\epsilon_{\mu\nu\sigma\rho}(\partial^{\mu}A^{\nu})
(\partial^{\sigma}K^{\dagger}){K}^{*\rho} 
+ ({\rm h.c.}),\nn\\ 
\label{gkv}
\mathcal{L}_{K^{*}N\Theta}&=&g_{K^{*}N\Theta}\overline{\Theta}
\gamma^{\mu}\overline{\Gamma}_{5}{K}^{*\dagger}_{\mu}N
+ ({\rm h.c.}). 
\label{vnt}
\eee
We neglect the tensor coupling of the $K^{*}N\Theta$ vertex for the lack
of information.  In order to determine the coupling constant
$g_{\gamma K K^{*}}$, we use the experimental data for the radiative
decay, which gives $0.388\, {\rm GeV}^{-1}$ for the neutral decay and 
$0.254 \,{\rm GeV}^{-1}$ for the charged
decay~\cite{particle,Liu:2003zi,Oh:2003kw}. 
$\overline{\Gamma}_{5}$ denotes ${\bf 1}_{4\times 4}$ for the
$\Theta^{+}_{+}$ and $\gamma_{5}$ for the $\Theta^{+}_{-}$.  Since we
have no information on $g_{K^{*}N\Theta}$ experimentally, we speculate
its value as $g_{K^{*}N\Theta}/g_{KN\Theta} = \pm 0.5$, assuming the ratio
similar to $g_{K^{*}N\Lambda}/g_{KN\Lambda}$.  Note that in
Refs.~\cite{Liu:2003zi,Yu:2003eq} the ratio of the couplings was taken to be 
$0.6$.  
In addition to $K^*$ exchange, we also consider $K_{1}(1270)$
axial-vector meson exchange.  However, since we find that its
contribution is tiny as in Ref.~\cite{Yu:2003eq}, we will not take
into account it in this work.   
Since the anomalous magnetic moment of $\Theta^{+}$ has not been fixed
experimentally, we need to rely on the model
calculations~\cite{Nam:2003uf,Zhao:2003gs,Kim:2003ay,Huang:2003bu,Liu:2003ab}.
Many of these calculations indicate small numbers for the $\Theta^+$ 
magnetic moment and hence negative values for the 
anomalous magnetic moment.  
As a typical value, 
we shall use for the anomalous magnetic moment 
$\kappa_{\Theta} = -0.8 \mu_N$.    

Now, we are in a position to calculate the invariant amplitudes for
the photo-production of the $\Theta^+$.  The amplitudes for $\gamma
n\rightarrow K^{-}\Theta^{+}$ in the PS scheme can be obtained as follows: 
\small     
\bee
i\mathcal{M}_{\rm s}
&=& 
eg\frac{\kappa_{n}}{4M_{n}}\overline{u}(p')\Gamma_{5}
\frac{F_{s}(\Slash{p}+\Slash{k}+M_{n})}
{(p+k)^{2}-M^{2}_{n}}
(\Slash{\epsilon}\Slash{k}-\Slash{k}\Slash{\epsilon})u(p),\nn\\
i\mathcal{M}_{\rm u}
&=&
-eg\overline{u}(p')\left(\Slash{\epsilon}
\frac{F^{n}_{c}(\Slash{p'}+M_{\Theta})-F_{u}\Slash{k}}
{(p'-k)^{2}-M^{2}_{\Theta}}
\Gamma_{5}-\frac{\kappa_{\Theta}}{4M_{\Theta}}
\overline{u}(\Slash{\epsilon}\Slash{k}-\Slash{k}\Slash{\epsilon})
\frac{F_{u}(\Slash{p'}-\Slash{k}+M_{\Theta})}
{(p'-k)^{2}-M^{2}_{\Theta}}\Gamma_{5}\right)u(p),\nn\\
i\mathcal{M}_{\rm t}
&=&
eg\overline{u}(p')\Gamma_{5}
\frac{F^{n}_{c}}{(k-k')^{2}-m^{2}_{K^{+}}}
u(p)(2k'\cdot\epsilon-k\cdot \epsilon),
\label{m1}
\eee
\normalsize
while that for the proton is derived as 
\small 
\bee
i\mathcal{M}_{\rm s}
&=&
-eg\overline{u}(p')\left(\Gamma_{5}
\frac{F^{p}_{c}(\Slash{p}+M_{p})+F_{s}\Slash{k}}
{(p+k)^{2}-M^{2}_{p}}
\Slash{\epsilon}-\frac{\kappa_{p}}{4M_{p}}
\Gamma_{5}
\frac{F_{s}(\Slash{p}+\Slash{k}+M_{p})}
{(p+k)^{2}-M^{2}_{p}}(\Slash{\epsilon}\Slash{k}-\Slash{k}\Slash{\epsilon})\right)u(p),\nn\\ 
i\mathcal{M}_{\rm u}
&=&
-eg\overline{u}(p')\left(\Slash{\epsilon}
\frac{F^{p}_{c}(\Slash{p'}+M_{\Theta})-F_{u}\Slash{k}}
{(p'-k)^{2}-M^{2}_{\Theta}}
\Gamma_{5}-\frac{\kappa_{\Theta}}{4M_{\Theta}}(\Slash{\epsilon}\Slash{k}-\Slash{k}\Slash{\epsilon}) 
\frac{F_{u}(\Slash{p'}-\Slash{k}+M_{\Theta})}
{(p'-k)^{2}-M^{2}_{\Theta}}\Gamma_{5}\right)u(p),
\label{m2}
\eee
\normalsize
where $\overline{u}$ and $u$ are the Dirac spinors of $\Theta^+$ and the
the nucleon. 
The four momenta $p$, $p'$, $k$ and $k'$ are for  
the nucleon, $\Theta^{+}$, photon, and the kaon, respectively.
Subscripts $\rm s$, $\rm u$ and $\rm t$  
stand for the Mandelstam variables. Note that in the case of the process
$\gamma p \rightarrow \overline{K}^0\Theta^+$, there is no
contribution from the meson-exchange diagram in the $t$--channel.  We have introduced the
form factors $F_{s,u,t}$ and $F^{n}_{c}$ 
in such a way that they satisfy the gauge
invariance~\cite{Ohta:ji,Haberzettl:1998eq,Davidson:2001qs} in the
form of  
\bee
F_{\rm \xi}=\frac{\Lambda^{4}}{\Lambda^{4}+\left({\rm \xi}-M^{2}_{\rm \xi}\right)^{2}},
\label{ff}
\eee
where $\xi$ represents relevant kinematic channels, $s$, $t$, and $u$,
generically.  The common form factor $F_{\rm c}$ is introduced 
according to the prescription suggested by
Refs.~\cite{Davidson:2001qs}:    
\bee
F^{\rm n}_{\rm c}&=&F_{\rm u}+F_{\rm t}-F_{\rm u}F_{\rm t},\nn\\F^{\rm
p}_{\rm c}&=&F_{\rm s}+F_{\rm u}-F_{\rm s}F_{\rm u}. 
\eee

In the PV scheme, we need to consider an additional contribution,
{\em i.e.}, the contact term, also known as the Kroll-Rudermann (KR)
term corresponding to diagram (d) in Fig.~\ref{dia}.  The term can 
be written as follows:
\be
i\mathcal{M}_{\rm KR,}=
-e\frac{g^{\ast}_{A}}{2f_{\pi}}\overline{u}(p')\Gamma_{5}\Slash{\epsilon}u(p)
\label{M11}.
\ee
While Yu {\em et al.}~\cite{Yu:2003eq} introduced the form
factors into the KR term in such a way that they satisfy the gauge
invariance, we make use of the following relation:
\bee
i\Delta\mathcal{M}^{0}=i\mathcal{M}^{0}_{\rm PV}-i\mathcal{M}^{0}_{\rm PS}&=& 
e\frac{g}{M_{N}+M_{\Theta}}
\left(\frac{\kappa_{\Theta}}{2M_{\Theta}}+\frac{\kappa_{N}}{2M_{N}}
\right) 
\overline{u}(p')\Gamma_{5}\Slash{\epsilon}\Slash{k}u(p).
\eee
Here, The superscript $0$ denotes the bare amplitudes without the form
factor.  
Since $i\Delta\mathcal{M}^{0}$ is gauge-invariant due to 
its tensor structure, 
we can easily insert the form factors, keeping the gauge invariance.  
Thus,  we arrive at the gauge-invariant amplitudes in 
the PV scheme as follows:
\bee
i\mathcal{M}_{\rm PV}&=&i\mathcal{M}_{\rm PS}+i\Delta\mathcal{M}\nn\\
&=&i\mathcal{M}_{\rm PS}+e\frac{g}{M_{N}+M_{\Theta}}
\left(F_{u}\frac{\kappa_{\Theta}}{2M_{\Theta}}+F_{s}\frac{\kappa_{N}}{2M_{N}}
\right) 
\overline{u}(p')\Gamma_{5}\Slash{\epsilon}\Slash{k}u(p).
\label{eq:gauge}
\eee    
Finally, the $K^{*}$-exchange amplitude is derived as 
follows:
\bee
\mathcal{M}_{K^{*}}=i\frac{F_{t}g_{\gamma K
K^{*}}g_{K^{*} N\Theta}}{(k-k')^{2}-M^{2}_{K^{*}}}\overline{u}(p')
\epsilon_{\mu\nu\sigma\rho}k^{\mu}\epsilon^{\nu}k'^{\sigma}\gamma^{\rho}
\overline{\Gamma}_{5}u(p), 
\eee 
which is clearly gauge-invariant.

\section{Numerical results}

\begin{figure}[tbh]
\begin{tabular}{cc}
\resizebox{8cm}{5cm}{\includegraphics{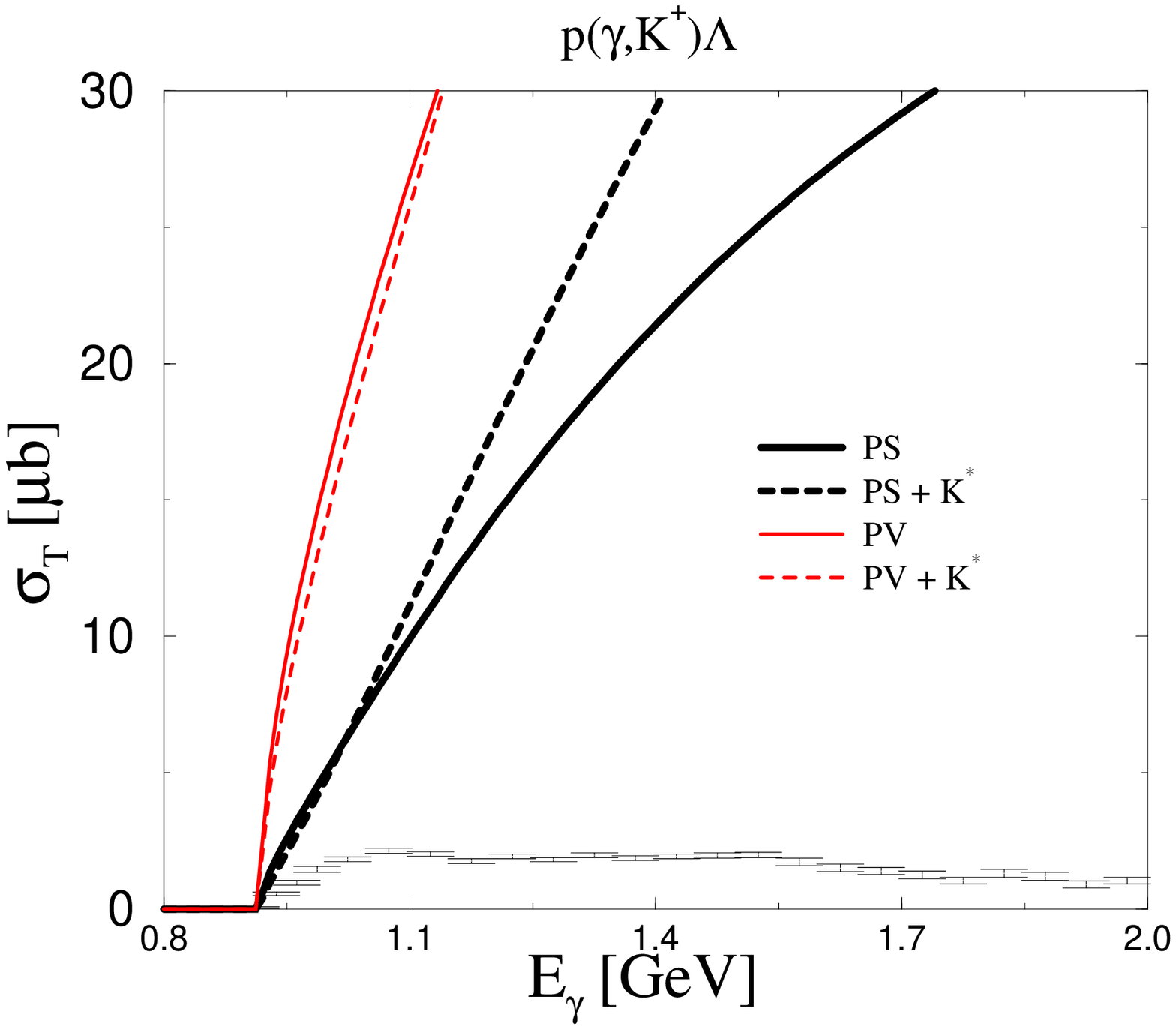}}
\resizebox{8cm}{5cm}{\includegraphics{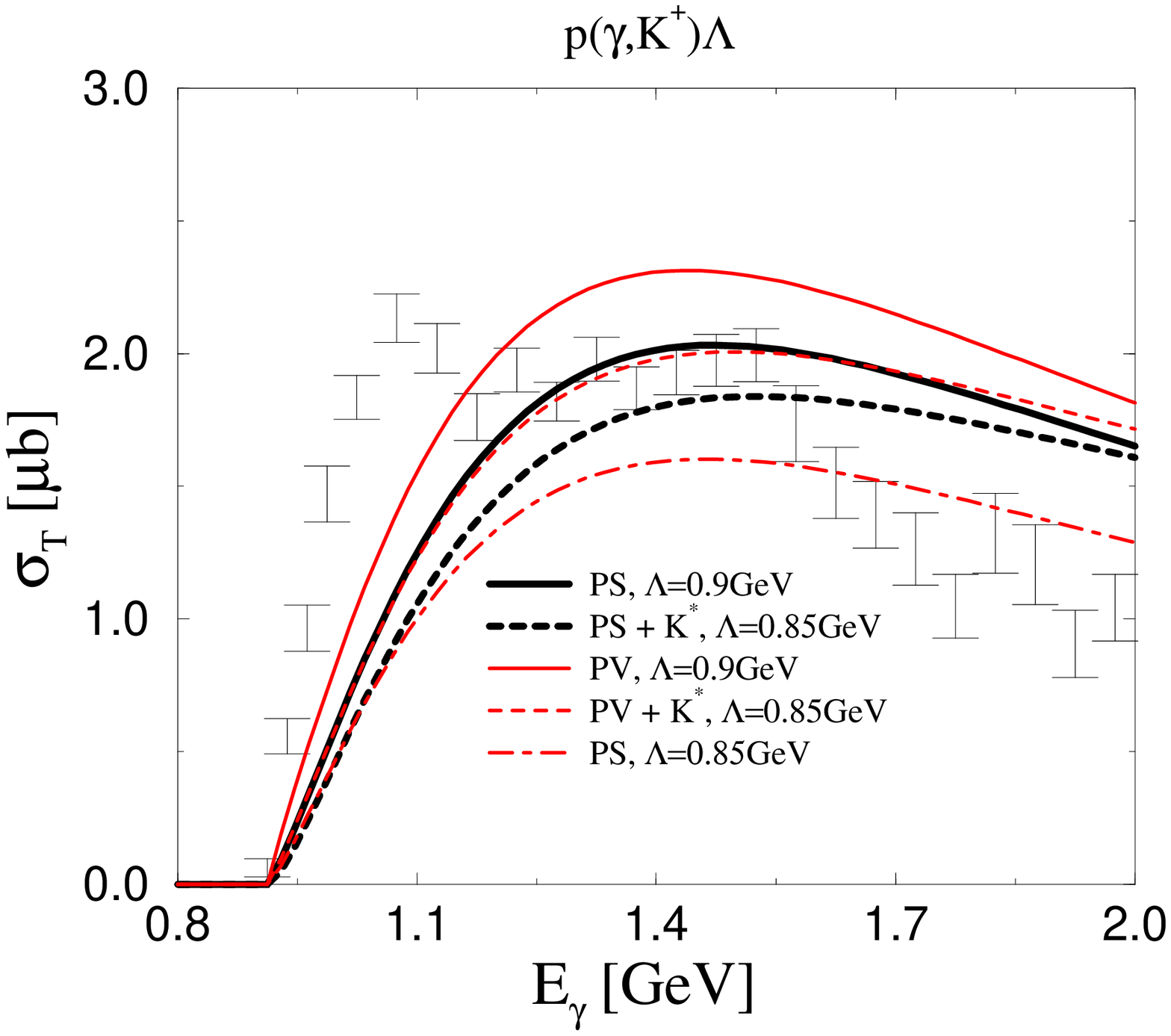}}
\end{tabular}
\caption{The total cross sections of
$\gamma p\rightarrow K^{+}\Lambda$ without (the left) and with (the right) the form
factors written in Eq.~(\ref{ff}). The experimental data was taken from 
Ref.~\cite{Tran:qw}.}
\label{nmset1}
\end{figure}

Before we calculate the photoproduction of the $\Theta^+$ numerically,
we need to fix the cutoff parameters in the form factors.  In doing
so, we will try to estimate the value of the cutoff parameters by
considering the process $\gamma p\rightarrow K^{+}\Lambda$, which is
known experimentally~\cite{Tran:qw} and the comparison of the theoretical 
prediction with the corresponding data is possible.  We have
calculated the Born diagrams as shown in Fig.~\ref{dia} for the
$K^{+}\Lambda$ production.  In Fig.~\ref{nmset1} we
present the total cross sections of the $\gamma p\rightarrow
K^{+}\Lambda$ reaction without the form factors.  Here, we have
employed the coupling constants $g_{KN\Lambda} = -13.3$ and
$g_{K^{*}N\Lambda} = -6.65$.  While the results without form factors
are monotonically increased unphysically as shown in the left panel of
Fig.~\ref{nmset1}, those with the form factors defined in
Eq.~(\ref{ff}) describe relatively well the experimental data as in
the right panel of Fig.~\ref{nmset1}.  We find that $\Lambda = 0.85
\sim 0.9\,{\rm GeV}$ give reasonable results qualitatively.  Note that
the peaks at around 1.0 GeV and 1.5 GeV in the experimental data are
believed to be related to higher nucleon resonances such as
$S_{11}(1650)$, $P_{11}(1710)$, $P_{13}(1720)$ and
$D_{13}(1895)$~\cite{Janssen:2001wk}, which in our calculations are
not included.      
\begin{figure}[tbh]
\begin{tabular}{cc}
\resizebox{8cm}{5cm}{\includegraphics{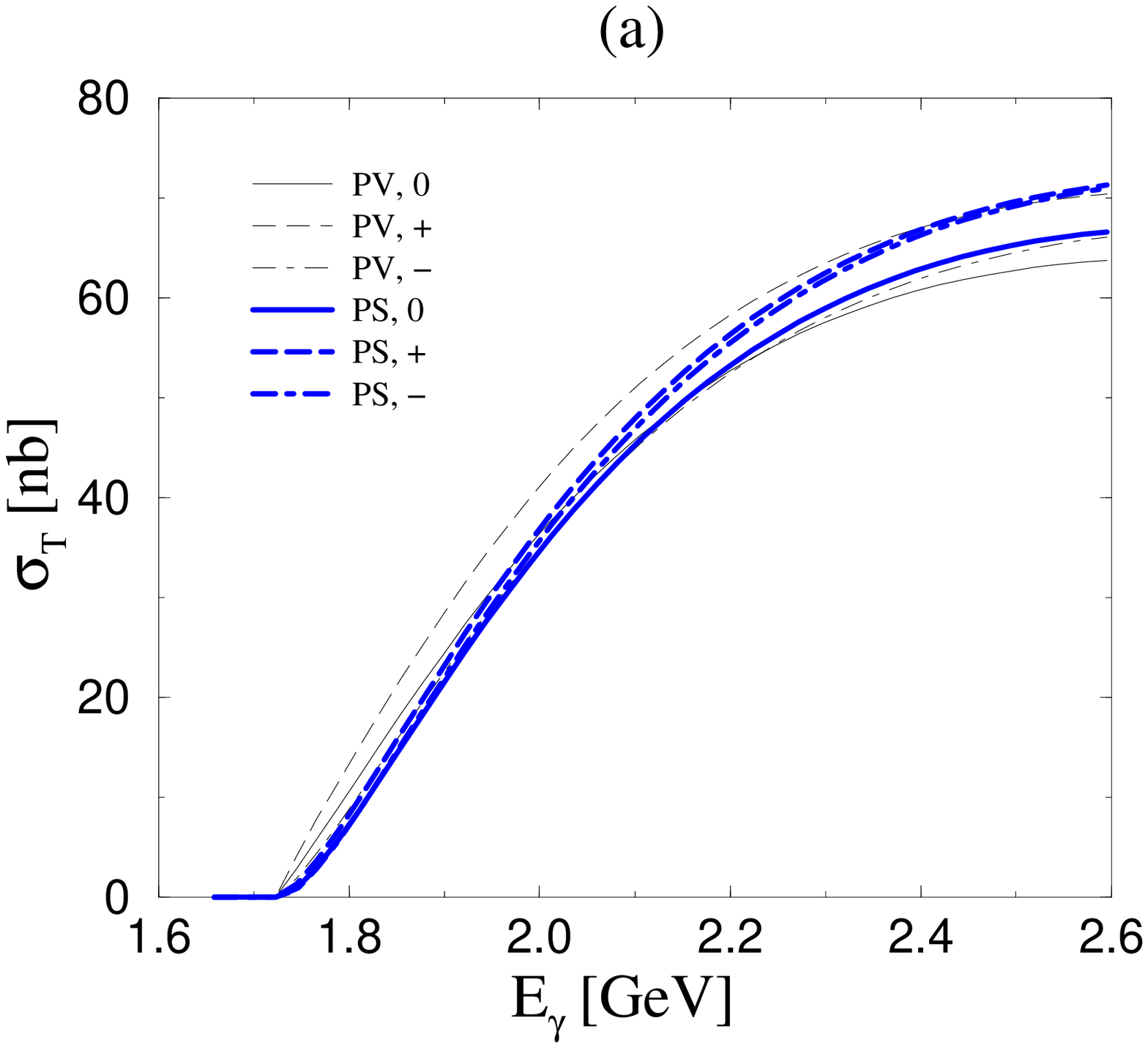}}
\resizebox{8cm}{5cm}{\includegraphics{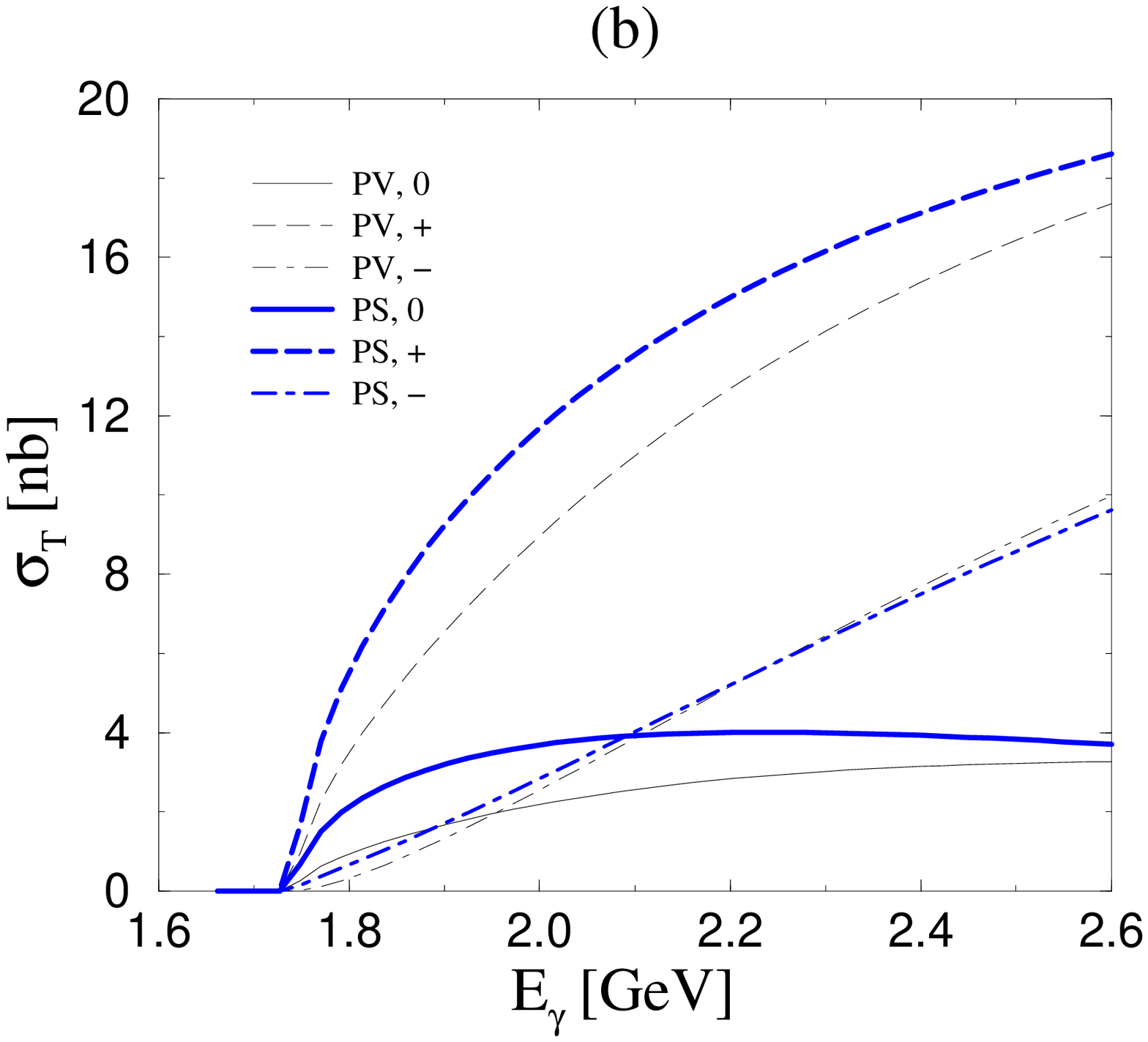}}
\end{tabular}
\caption{The total cross sections for the reactions of $\gamma
n\rightarrow K^{-}\Theta^+_{+} $
(a) and $\gamma p\rightarrow  \overline{K}^{0}\Theta^+_{+}$ (b). PV and PS indicate the 
coupling schemes. 0, + and - indicate
$g_{K^{*}N\Theta}=0$, $g_{K^{*}N\Theta}=g_{KN\Theta}/2$ and
$g_{K^{*}N\Theta}=-g_{KN\Theta}/2$, respectively.}
\label{nmset2}
\end{figure}    

Based on these results, we assume that the cutoff parameter for the
$KN\Theta$ vertex is the same as for the $KN\Lambda$ one
and use $\Lambda = 0.85$ GeV.  Figure~\ref{nmset2} draws the total
cross sections with the form factors and $g_{K^{*}N\Theta}$ being varied
between $-g_{KN\Theta}/2$ and $g_{KN\Theta}/2$.  
We see that the
differences between the PV and PS schemes turn out to be small,
as compared to the results of Ref.~\cite{Yu:2003eq}.  
The reason lies in the fact that 
Ref.~\cite{Yu:2003eq} introduced the form factor in the KR term
directly, while we employ the relation between the PV and PS
schemes as given in Eq.(\ref{eq:gauge}).  It is very natural that in 
the low-energy limit the difference between the PV and PS schemes should
disappear.  In this sense, the present results is consistent 
with the low-energy relation for the photo-production.  

Coming to the photo-production of the $\Theta^+$ in the $\gamma
p\rightarrow \overline{K}^0 \Theta^+$ reaction, we notice that the
total cross section is smaller than the case of $\gamma n$ and 
rather sensitive to the contribution of $K^*$ exchange.  It can be
understood by the fact that the contribution of $K$ exchange is absent
and the $s$-- and $u$--channels are suppressed by the form factors.  The
average values of the total cross sections are estimated as 
follows: $\sigma_{\gamma n\rightarrow K^-\Theta^+}\sim 44 \,{\rm nb}$
and $\sigma_{\gamma p\rightarrow \overline{K}^0\Theta^+}\sim 13 \,{\rm
  nb}$ in the range of the photon energy $1.73\,{\rm GeV} < E_{\gamma} <
2.6 \,{\rm GeV}$.  Note that these values are smaller than those of
Ref.~\cite{Nam:2003uf}, where $\Lambda = 1.0\,{\rm  GeV}$ is employed.  
\begin{figure}[tbh]
\begin{tabular}{cc}
\resizebox{8cm}{5cm}{\includegraphics{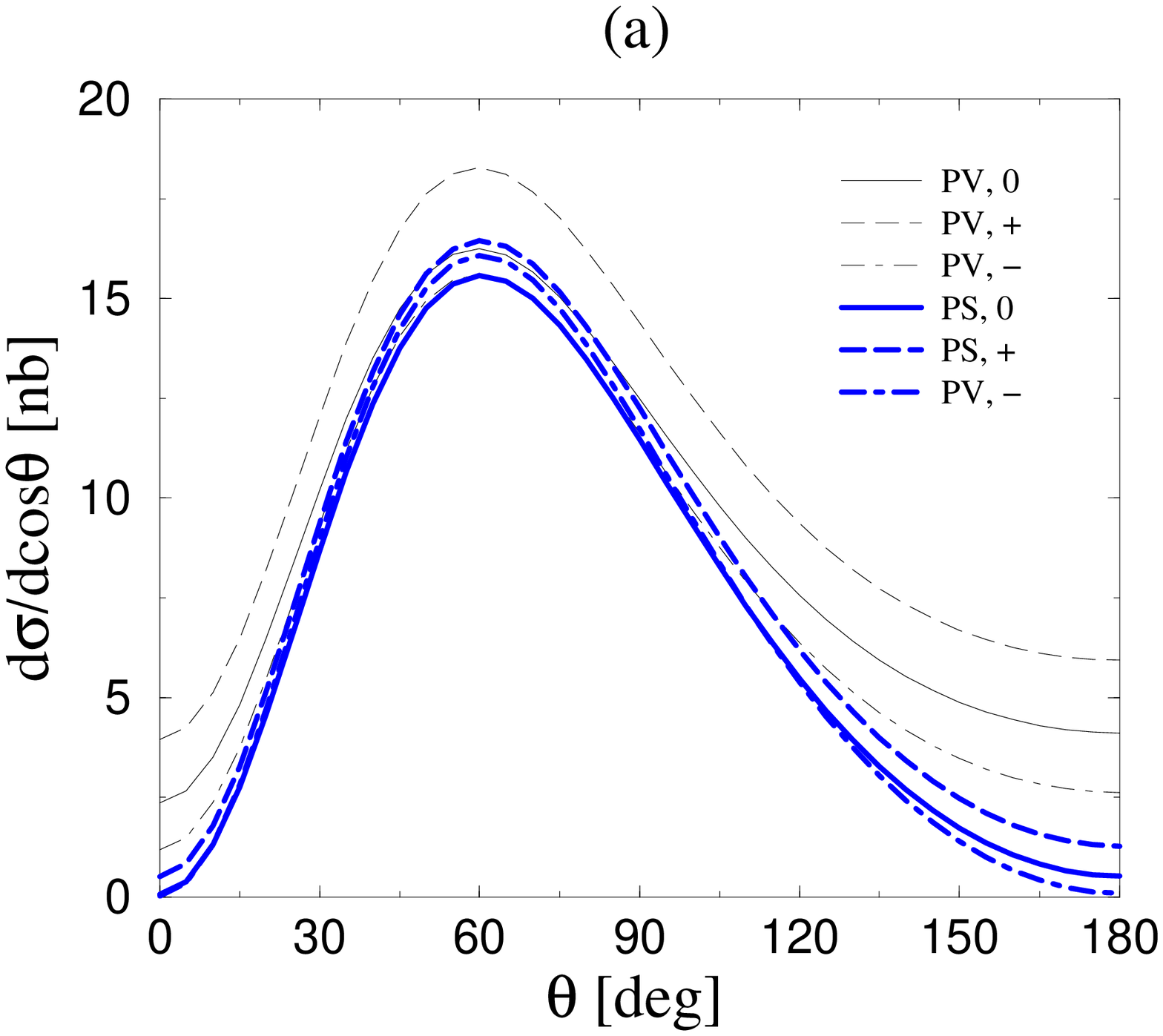}}
\resizebox{8cm}{5cm}{\includegraphics{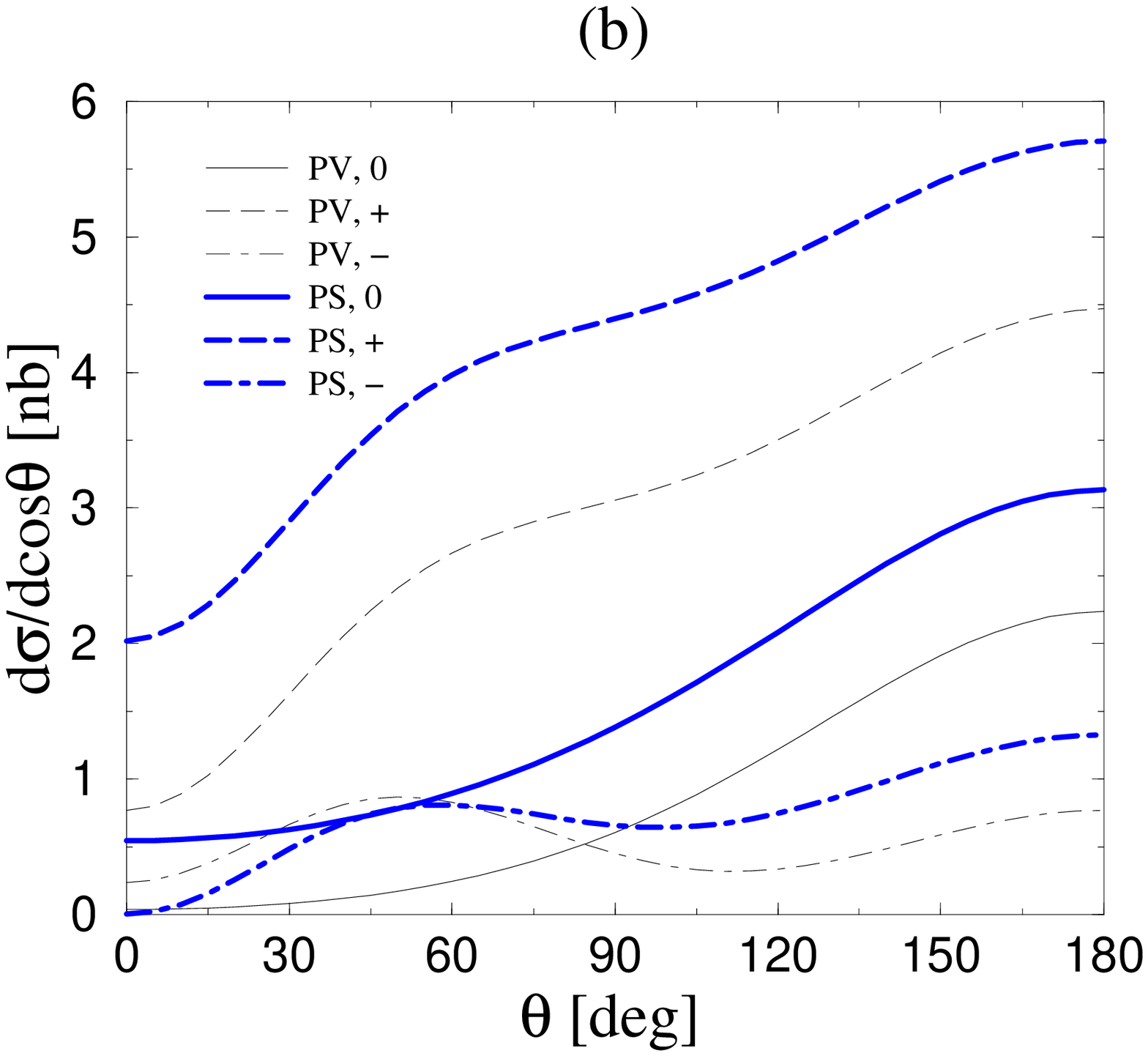}}
\end{tabular}
\caption{The differential cross sections for the reactions of $\gamma
n\rightarrow K^{-}\Theta^{+}_{+}$ (a) and $\gamma
p\rightarrow  \overline{K}^{0}\Theta^{+}_{+}$ (b) at $\sqrt{s} = 2.1\,{\rm  GeV}$.}
\label{nmset4}
\end{figure} 

In Fig.~\ref{nmset4}, we draw the differential cross sections.  In the
case of the $\gamma n\rightarrow K^-\Theta^+$, 
the peak around $60^\circ$ is clearly seen as shown in the left panel
of Fig.~\ref{nmset4}.   
This peak is caused by the $t$--channel dominance which brings about  
the combination of the factor
$|\epsilon\cdot k^{'}|^{2}\sim \sin^{2}\theta$ and the form factor.  
In the multipole basis, an $M1$ amplitude is responsible for it.  
In contrast, for the production from the proton, 
$K$ exchange is absent, and the role of $K^{*}$ exchange and its
interference with the $s$-- and $u$--channel diagrams become more 
important.  Therefore, the differential cross section of the  
$\gamma p\rightarrow \overline{K}^0\Theta^+$ process is quite
different from that of the $\gamma n\rightarrow K^-\Theta^+$.  The
present results look rather different from those of 
Ref.~\cite{Oh:2003kw}, where the relation $g_{K^{*}N\Theta}=\pm
g_{KN\Theta}$ was employed.  It is so since the amplitude of $K^*$ 
exchange is twice as large as that in the present work, 
and has an even more important contribution to the amplitudes.  We
need more experimental information in order to settle the uncertainty
in the reaction mechanism.   

\begin{figure}[tbh]
\begin{tabular}{cc}
\resizebox{8cm}{5cm}{\includegraphics{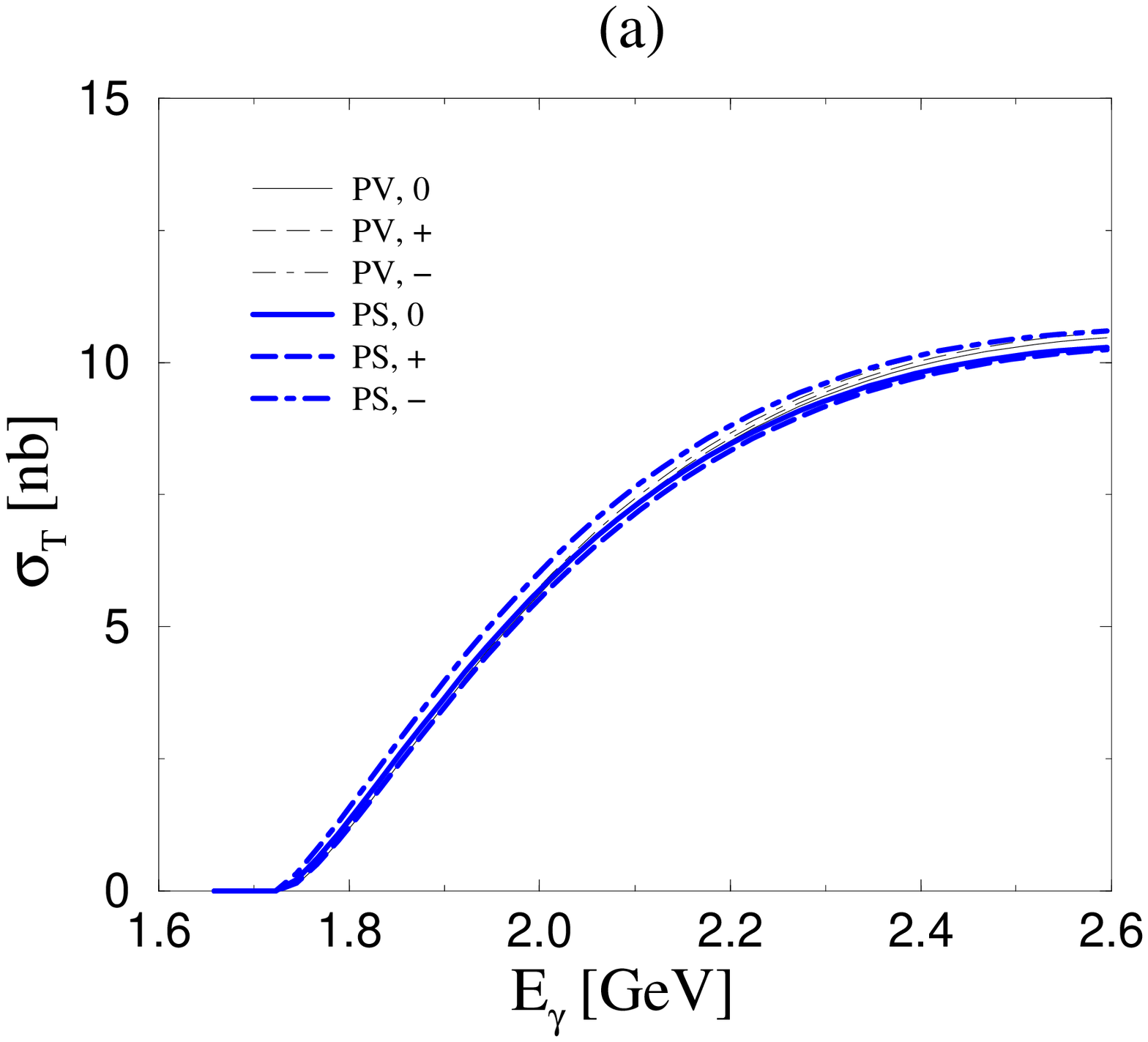}}
\resizebox{8cm}{5cm}{\includegraphics{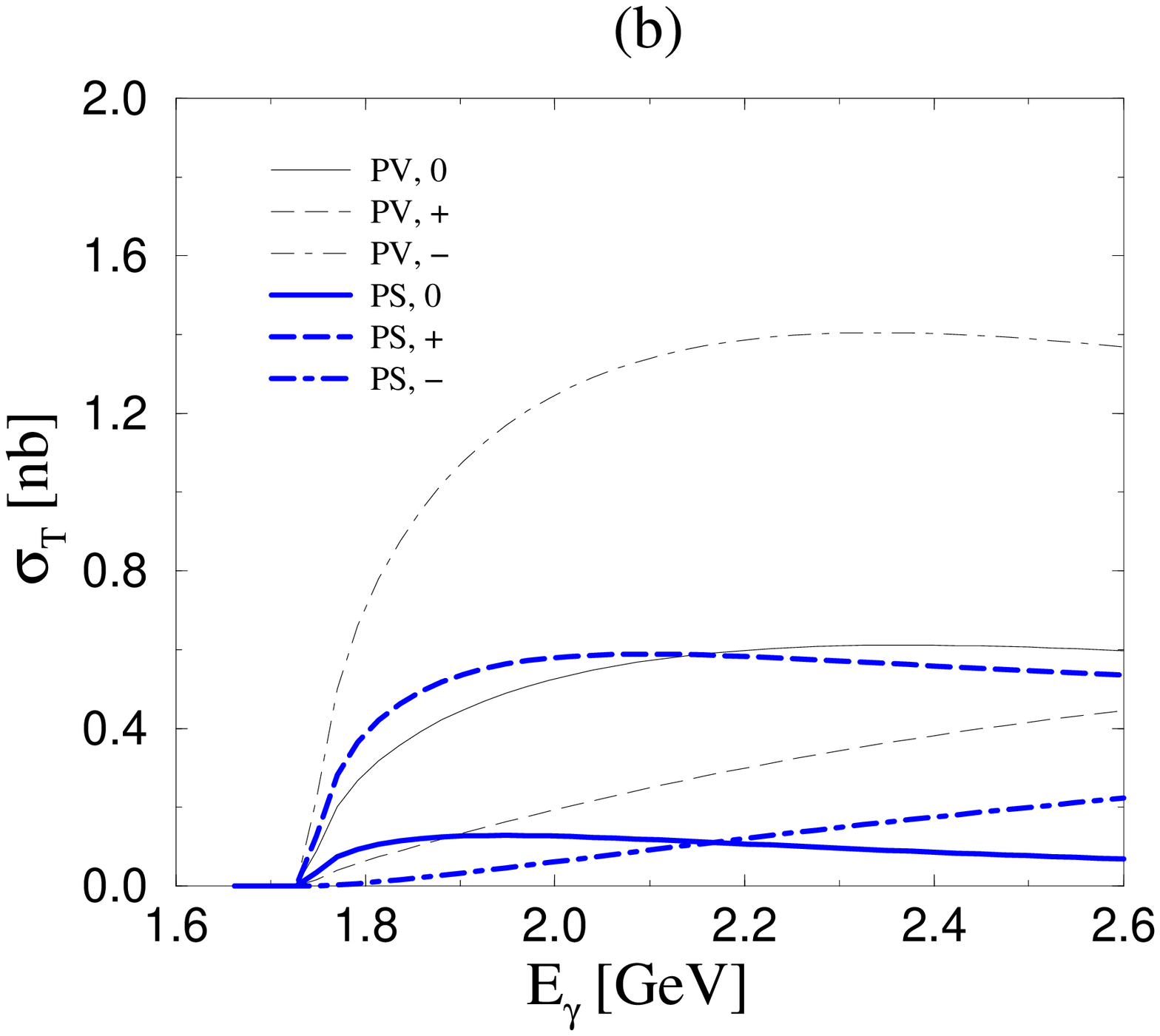}}
\end{tabular}
\caption{The total cross sections for the reactions of $\gamma
n\rightarrow K^{-}\Theta^+_{-}$ (a) and 
$\gamma p\rightarrow \overline{K}^{0}\Theta^+_{-}$ (b).}
\label{nmset5}
\end{figure} 

We now present the total cross sections for the negative parity 
$\Theta^{+}_{-}$ in Fig.~\ref{nmset5}.  
The contribution of $K^*$ exchange is almost
negligible in the case of the $\gamma n\rightarrow K^-\Theta^+$
process, whereas it plays a main role in 
$\gamma p\rightarrow \overline{K}^0\Theta^+$.  The total cross
sections for the negative-parity $\Theta^+$ turn out to be
approximately ten times smaller than those for the positive-parity
one.  This fact pervades rather universally various reactions for
the $\Theta^+$ production.  The reason is that the momentum-dependent
$p$-wave coupling $\vec \sigma \cdot \vec q$ for the positive parity
$\Theta^+$ enhances the coupling strength effectively at the momentum
transfer $|\vec q| \sim 1$ GeV, a typical value for the $\Theta^+$
production using non-strange particles.  The enhancement factor is
about 1 GeV/0.26 GeV, where 0.26 GeV is the kaon momentum in the  
$\Theta^+$ decay.  Therefore, the cross sections become larger for the
positive parity case than for the negative parity case by a factor 
$(1/0.26)^2 \sim 10$.  

\begin{figure}[tbh]
\begin{tabular}{cc}
\resizebox{8cm}{5cm}{\includegraphics{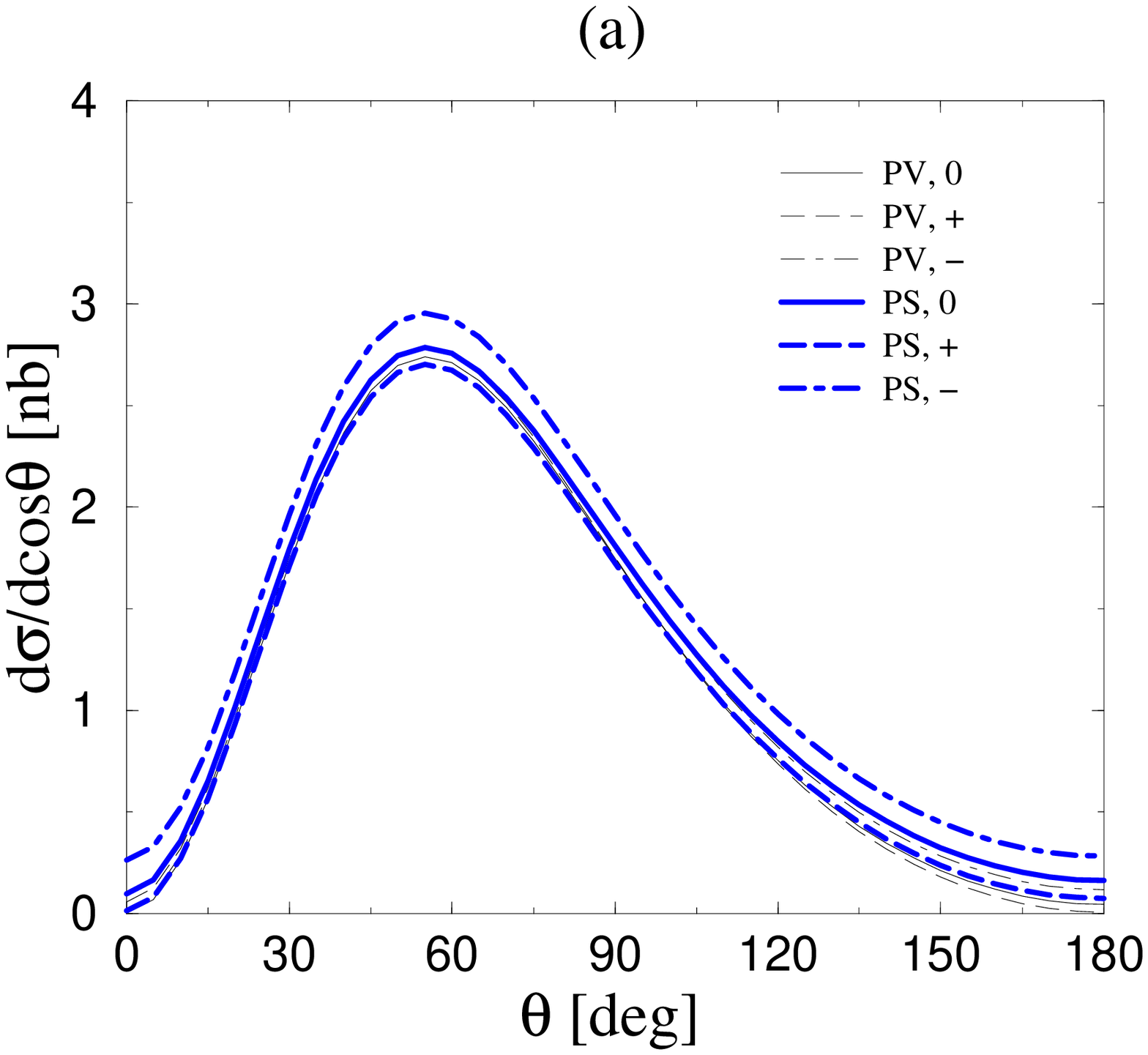}}
\resizebox{8cm}{5cm}{\includegraphics{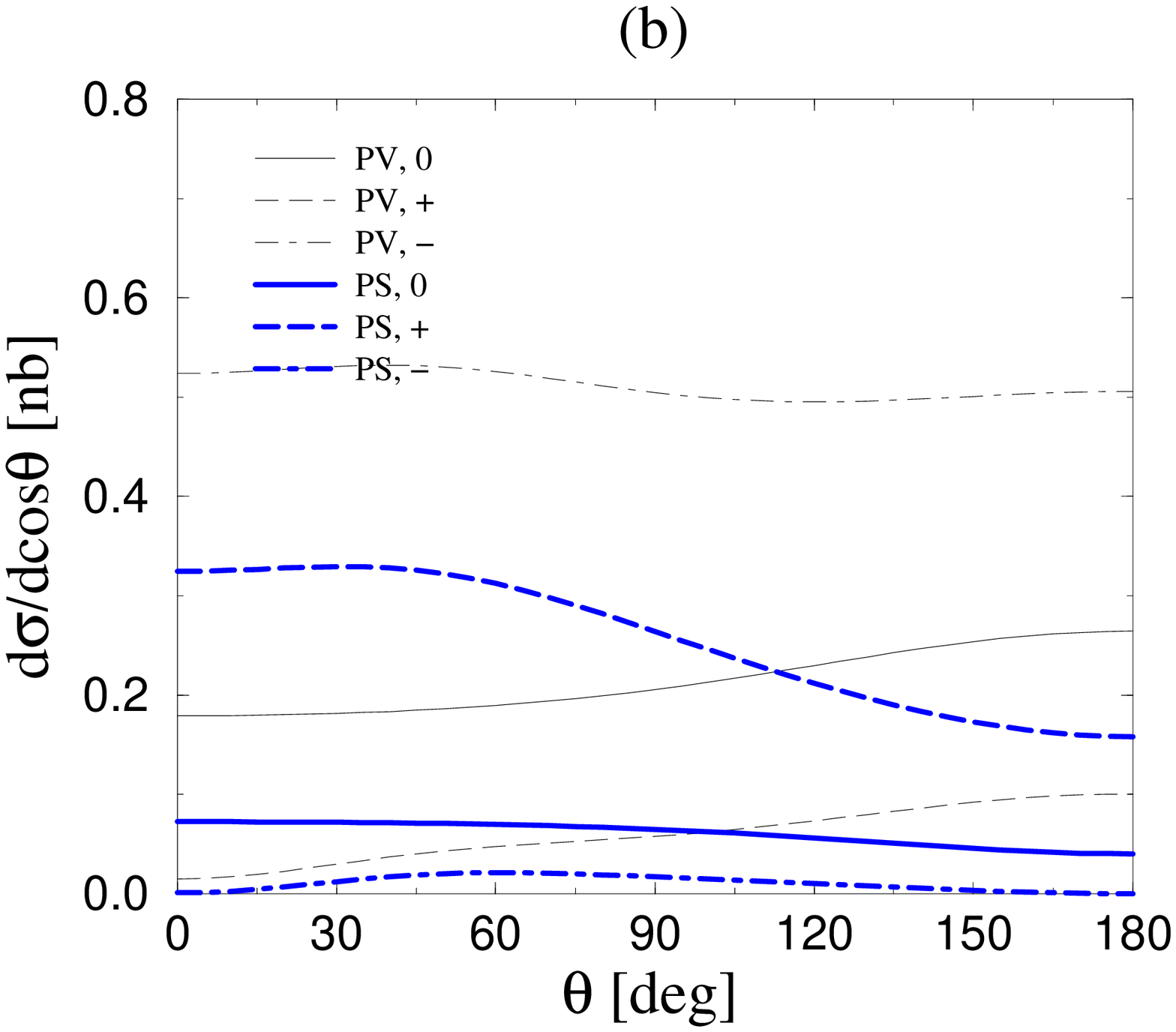}}
\end{tabular}
\caption{The differential cross sections for the reactions of $\gamma
n\rightarrow K^{-}\Theta^+_{-}$ (a) and 
$\gamma p\rightarrow \overline{K}^{0}\Theta^+_{-}$ (b) at $\sqrt{s} = 2.1\,{\rm  GeV}$.} 
\label{nmset6}
\end{figure} 

The differential cross sections for the $\Theta^{+}_{-}$ photo-production are drawn in
Fig.~\ref{nmset6}.  The peak around $60^\circ$ appears in the $\gamma
n$ interaction as in the case of the $\Theta^{+}_{+}$.  That for the
production via the $\gamma p$ interaction shows quite different from
the case of the $\Theta_{+}^{+}$.  
  
\section{Summary and Discussions}
We investigated $\gamma N \rightarrow \overline{K}\Theta^{+}$ reactions with
the Born approximation including $t$--channel $K^*$ exchange.  
In order to make our discussions quantitative, 
we employed the phenomenological
strong form factor with the cutoff, $\Lambda$, which was determined by
$p(\gamma,K^{+})\Lambda$ reaction without $K_{1}$ and the 
nucleon resonances.  Then we obtained $\Lambda$ = 0.85 $\sim$ 0.9 GeV
with about $30\%$ tolerance and took $\Lambda$ = 0.85 GeV 
for the numerical calculations.  We also treated them in the pseudoscalar
(PS) and pseudovector (PV) coupling schemes.  Then we constructed the
gauge-invariant amplitudes in the PS and PV using the
relation, $i\mathcal{M}_{PV}=i\mathcal{M}_{PS}+i\Delta\mathcal{M}$. 
In this method, the result in the PV becomes rather similar to that of 
the PS as expected from the low-energy limit.  This behaviour is
deeply related to the prescription of the form factor which we
employed. As shown in Fig.~\ref{nmset2}, our form factor suppressed
the $u$-- and $s$--channels more than that of the $t$--channel.
However, this situation is not accidental but explains the physics
about the amplitude, which extracts the most dominant one from the
Born amplitudes in the kinematical channels: The intermediate 
states in the s- and $u$--channels (N and $\Theta^+$) are further off-shell 
than that in the $t$--channel (K).  Reminding of that
$\kappa_{\Theta}$ is only contained in $s$-- and $u$--channel
amplitudes, it is natural for us to have the cross sections which are
not dependent much on $\kappa_{\Theta}$. Consequently in this method,
we were able to diminish the model and parameter dependences. We note
that these results are rather different from those of 
Ref.~\cite{Yu:2003eq} in which the authors modified Kroll-Ruderman
term with the form factors directly in order to keep the gauge
invariance. However, as for the PS scheme 
only, their results are essentially equivalent to ours. 

For the total cross sections, we found that 
$\sigma_{\gamma n \rightarrow K^{-}\Theta^+} (44 \; {\rm nb}) 
>  
\sigma_{\gamma p\rightarrow \overline{K}^{0}\Theta^+} (13 \; {\rm nb})$ 
for the positive parity $\Theta^{+}_{+}$.  
This was a similar result as obtained in Refs.~\cite{Liu:2003rh,Liu:2003zi}, 
in which they used one overall form factor and ignored the anomalous 
magnetic moments ($\kappa_{\Theta} = \kappa_N = 0$).  
Once again in our calculations the form factor played an important role here.  
In Ref.~\cite{Oh:2003kw}, they obtained the cross sections for 
$\gamma n$ and $\gamma p$ processes similar by employing a larger 
value of the $K^*N\Theta$ coupling than ours 
($g_{K^* N \Theta} = g_{K^* N \Theta}$).  
This value produced the total cross sections consistent with the data
from SAPHIR.  
However, more experimental analyses should be necessary in order to 
confirm the absolute value of the total cross section.  
In Ref.~\cite{Yu:2003eq}, they also obtained similar total cross sections
both for $\gamma n$ and $\gamma p$ reactions by employing a large
cutoff parameter $\Lambda = 1.8$ GeV.  

So far, we have variable theoretical predictions based on different 
reaction mechanisms and model parameters.  
More experimental information will be necessary in order to pin down 
such uncertain situation.  
However, it is a universal feature that 
the total cross section for the positive parity 
$\Theta^+$ production is about factor ten larger than that of the 
negative parity one.  
This might be useful when proceeding step by step to obtain 
more information about the nature of the pentaquark baryon $\Theta^+$.

\section*{Acknowledgment}
We thank Hiroshi Toki and 
Takashi Nakano for discussions and comments. 
The work of H.-Ch.Kim is supported by the Korean Research Foundation
(KRF--2003--070--C00015) and in part by
the 21st COE Program ``Towards A New Basic
Science: Depth and Synthesis" (Osaka university). 
The work of S.I.Nam has been supported by 
the scholarship endowed from the Ministry of Education, Science,
Sports and Culture of Japan. 
A.H. would like to thank the hospitality of the members of the NuRI 
at Pusan National University.  



\begin{thebibliography}{99}
\bibitem{Nakano:2003qx}
T.~Nakano {\it et al.}  [LEPS Collaboration],
Phys.\ Rev.\ Lett.\  {\bf 91}, 012002 (2003)
\bibitem{Diakonov:1997mm}
D.~Diakonov, V.~Petrov and M.~V.~Polyakov,
Z.\ Phys.\ A {\bf 359}, 305 (1997) 
\bibitem{Barmin:2003vv}
V.~V.~Barmin {\it et al.}  [DIANA Collaboration],
Phys.\ Atom.\ Nucl.\  {\bf 66}, 1715 (2003)
[Yad.\ Fiz.\  {\bf 66}, 1763 (2003)]
\bibitem{Stepanyan:2003qr}
S.~Stepanyan {\it et al.}  [CLAS Collaboration],
Phys.\ Rev.\ Lett.\  {\bf 91}, 252001 (2003) 
\bibitem{Kubarovsky:2003fi}
V.~Kubarovsky {\it et al.}  [CLAS Collaboration],
Erratum-ibid.\  {\bf 92}, 049902 (2004)
[Phys.\ Rev.\ Lett.\  {\bf 92}, 032001 (2004)]
\bibitem{Barth:2003es}
J.~Barth {\it et al.}  [SAPHIR Collaboration],
hep-ex/0307083
\bibitem{Airapetian:2003ri}
A.~Airapetian {\it et al.}  [HERMES Collaboration],
hep-ex/0312044
\bibitem{private1}
Private conversation with T~.Nakano 
\bibitem{Alt:2003vb}
C.~Alt {\it et al.}  [NA49 Collaboration],
Phys.\ Rev.\ Lett.\  {\bf 92}, 042003 (2004)
\bibitem{Praszalowicz:2003tc}
M.~Prasza\l owicz,
Phys.\ Lett.\ B {\bf 583}, 96 (2004)
\bibitem{Karliner:2004qw}
M.~Karliner and H.~J.~Lipkin,
arXiv:hep-ph/0401072.
\bibitem{Sugiyama:2003zk}
J.~Sugiyama, T.~Doi and M.~Oka,
Phys.\ Lett.\ B {\bf 581}, 167 (2004)
\bibitem{Zhu:2003ba}
S.~L.~Zhu,
Phys.\ Rev.\ Lett.\  {\bf 91}, 232002 (2003)
\bibitem{Sasaki:2003gi}
S.~Sasaki,
hep-lat/0310014
\bibitem{Csikor:2003ng}
F.~Csikor, Z.~Fodor, S.~D.~Katz and T.~G.~Kovacs,
JHEP {\bf 0311}, 070 (2003)
\bibitem{Stancu:2003if}
F.~Stancu and D.~O.~Riska,
Phys.\ Lett.\ B {\bf 575}, 242 (2003)
\bibitem{Glozman:2003sy}
L.~Y.~Glozman,
Phys.\ Lett.\ B {\bf 575}, 18 (2003)
\bibitem{Hosaka:2003jv}
A.~Hosaka,
Phys.\ Lett.\ B {\bf 571}, 55 (2003)

\bibitem{Huang:2003we}
F.~Huang, Z.~Y.~Zhang, Y.~W.~Yu and B.~S.~Zou,
hep-ph/0310040
\bibitem{Liu:2003rh}
W.~Liu and C.~M.~Ko,
Phys.\ Rev.\ C {\bf 68}, 045203 (2003)
\bibitem{Hyodo:2003th}
T.~Hyodo, A.~Hosaka and E.~Oset,
Phys.\ Lett.\ B {\bf 579}, 290 (2004)
\bibitem{Oh:2003kw}
Y.~S.~Oh, H.~C.~Kim and S.~H.~Lee,
Phys.\ Rev.\ D {\bf 69}, 014009 (2004)
\bibitem{nam2}
S.~I.~Nam, A.~Hosaka and H.-Ch.~Kim,
hep-ph/0402138
\bibitem{nam3}
S.~I.~Nam, A.~Hosaka and H.-Ch.~Kim, in preparation
\bibitem{Thomas:2003ak}
A.~W.~Thomas, K.~Hicks and A.~Hosaka,
hep-ph/0312083, to appear in Prog. Theor. Phys.
\bibitem{Hanhart:2003xp}
C.~Hanhart {\it et al.},
hep-ph/0312236
\bibitem{Liu:2003zi}
W.~Liu and C.~M.~Ko,
nucl-th/0309023
\bibitem{Nam:2003uf}
S.~I.~Nam, A.~Hosaka and H.-Ch.~Kim,
Phys.\ Lett.\ B {\bf 579}, 43 (2004)

\bibitem{Zhao:2003gs}
Q.~Zhao,
hep-ph/0310350.
\bibitem{Yu:2003eq}
B.~G.~Yu, T.~K.~Choi and C.~R.~Ji,
nucl-th/0312075.
\bibitem{Liu:2003ab}
Y.~R.~Liu, P.~Z.~Huang, W.~Z.~Deng, X.~L.~Chen and S.~L.~Zhu,
arXiv:hep-ph/0312074
\bibitem{Li:2003cb}
W.~W.~Li, Y.~R.~Liu, P.~Z.~Huang, W.~Z.~Deng, X.~L.~Chen and S.~L.~Zhu,
arXiv:hep-ph/0312362
\bibitem{Ko:2003xx}
P.~Ko, J.~Lee, T.~Lee and J.~h.~Park,
arXiv:hep-ph/0312147.
\bibitem{Kim:2003ay}
H.-~Ch.~Kim and M. Prasz\l owicz, to appear in Phys. Lett. B,  
hep-ph/0308242
\bibitem{Huang:2003bu}
P.~Z.~Huang, W.~Z.~Deng, X.~L.~Chen and S.~L.~Zhu,
hep-ph/0311108
\bibitem{Ohta:ji}
K.~Ohta,
Phys.\ Rev.\ C {\bf 40} (1989) 1335
\bibitem{Haberzettl:1998eq}
H.~Haberzettl, C.~Bennhold, T.~Mart and T.~Feuster,
 Phys.\ Rev.\ C {\bf 58} (1998) 40
\bibitem{Davidson:2001qs}
R.~M.~Davidson and R.~Workman,
nucl-th/0101066

\bibitem{Janssen:2001wk}
S.~Janssen, J.~Ryckebusch, D.~Debruyne and T.~Van Cauteren,
Phys.\ Rev.\ C {\bf 65}, 015201 (2002) 
\bibitem{particle}
Particle Data Group, K. Hagiwara {\it et al.}, Phys. Rev. D {\bf 66},
01001 (2002)
\bibitem{Cheoun:kn}
M.~K.~Cheoun, B.~S.~Han, I.~T.~Cheon and B.~G.~Yu,
Phys.\ Rev.\ C {\bf 54}, 1811 (1996)
\bibitem{Tran:qw}
M.~Q.~Tran {\it et al.}  [SAPHIR Collaboration],
Phys.\ Lett.\ B {\bf 445}, 20 (1998)
\bibitem{stokes}
V. G. J. Stokes and Th. A. Rijken, Phys. Rev. C {\bf 59}, 3009 (1999)



\end{thebibliography}
\end{document}